# The tetrazole analogue of the auxin indole-3-acetic acid binds preferentially to TIR1 and not AFB5.

**Mussa Quareshy** [1,†], **Justyna Prusinska** [1], **Martin Kieffer** [2], **Kosuke Fukui** [3], **Alonso J. Pardal** [1], **Silke Lehmann** [1,5], **Patrick Schafer** [1,5], **Charo I. del Genio** [1], **Stefan Kepinski** [2], **Kenichiro Hayashi** [3], **Andrew Marsh** [4], **Richard M. Napier** [1,†]

[1] School of Life Sciences, University of Warwick, Coventry, CV4 7AL, UK

[2] Centre for Plant Sciences, University of Leeds, Leeds LS2 9JT

[3] Department of Biochemistry, Okayama University of Science, 1-1 Ridaicho, Kita-ku, Okayama-shi Okayama, JP 700-0005

[4] Department of Chemistry, University of Warwick, Coventry, CV4 7AL, UK

[5] Warwick Integrative Synthetic Biology Centre

[†] Corresponding authors:

mussaquareshy@gmail.com; Richard.napier@warwick.ac.uk

**Highlights:** We apply a functional group swap to auxin yielding an active auxin that also confers selectivity between 2 members of the auxin receptor family.

**Who did what:** MQ came up with the concept of iMT, performed most of the experiments including synthesis of iMT in AM's lab with AM's guidance in designing the synthesis. JP performed the SPR SCK assays. MK performed the pull-down assays and SK provided the *tir1-1* knockout lines. KH and KF synthesised the 4,5 and 6 Cl iMT analogues as well as additional iMT, performed the *DR5::GUS* and *DII::VENUS* reporter assays. CIDG performed the curve fitting analysis to derive the primary root growth $IC_{50}$ values. AJP, SL and PS performed the protoplast reporter assays and AJP did the qPCR. RN helped to develop the concept of iMT and was involved in the design of lead optimisation. MQ and RN wrote the manuscript with contributions from all the other authors.

**Abstract**

Auxin is considered one of the cardinal hormones in plant growth and development. It regulates a wide range of processes throughout the plant. Synthetic auxins exploit the auxin-signalling pathway and are valuable as herbicidal agrochemicals. Currently, despite a diversity of chemical scaffolds all synthetic auxins have a carboxylic acid as the active core group. By applying bio-isosteric replacement we discovered that indole-3-tetrazole was active by surface plasmon resonance (SPR) spectrometry, showing that the tetrazole could initiate assembly of the TIR1 auxin co-receptor complex. We then tested the tetrazole's efficacy in a range of whole plant physiological assays and in protoplast reporter assays which all confirmed auxin activity, albeit rather weak. We then tested indole-3-tetrazole against the AFB5 homologue of TIR1, finding that binding was selective against TIR1, absent with AFB5. The kinetics of binding to TIR1 are contrasted to those for the herbicide picloram, which shows the opposite receptor preference as it binds to AFB5 with far greater affinity than to TIR1. The basis of the preference of indole-3-tetrazole for TIR1 was revealed to be a single residue substitution using molecular docking, and assays using *tir1* and *afb5* mutant lines confirmed selectivity *in vivo*. Given the potential that a TIR1-selective auxin might have for unmasking receptor-specific actions, we followed a rational design, lead optimisation campaign and a set of chlorinated indole-3-tetrazoles was synthesised. Improved affinity for TIR1 and the preference for binding to TIR1 was maintained for 4- and 6-chloroindole-3-tetrazoles, coupled with improved efficacy *in vivo*. This work expands the range of auxin chemistry for the design of receptor-selective synthetic auxins.

## Introduction

Auxin (indole-3-acetic acid; IAA (1) Figure 1) regulates diverse developmental processes including cell elongation, cell division, tropic responses, lateral rooting and branching, and synthetic auxins are an important class of selective herbicides. The best-studied auxin receptor is the F-box protein Transport Inhibitor Resistant 1 [1]. A wealth of experimentation has shown that TIR1 is the F-box component of a Skp, Cullin, F-box (SCF) type E3 ubiquitin ligase complex SCF[TIR1], although it wasn't until seminal work by two groups [2,3] that TIR1 was shown to be the major auxin receptor. The TIR1 family includes five additional Auxin F-Box proteins (AFBs) in *Arabidopsis* and activity within this family has been shown to be largely redundant [4], with some notable exceptions. Root architecture responses to nitrate levels appear to be mediated through AFB3 [5,6] and AFB5 has been shown to be the dominant site of action for the picolinate herbicides [7,8].

The mechanism of auxin action is coordinated through transcriptional regulation and this has been reviewed extensively [2,3,9,10,11]. At low auxin concentrations, AUX/IAA transcriptional repressor proteins, together with co-repressor (TOPLESS) proteins repress genes targeted by the Auxin Response Factor (ARF) transcriptional activators. As concentrations rise, auxin binds to TIR1 creating a high-affinity surface for recruitment of the AUX/IAA co-receptor. The assembly of this SCF[TIR1] co-receptor complex has been used for an auxin binding assay using surface plasmon resonance spectrometry [8,12]. Assembly of the complex *in vivo* leads to ubiquination of the AUX/IAAs and consequent degradation in the proteasome. The consequent reduction in concentrations of AUX/IAA proteins releases the ARFs, allowing transcription to commence. The TIR1 crystal structure [13] provided the paradigm for IAA acting as molecular glue between TIR1 and AUX/IAA, and radiolabel and SPR binding experiments illustrated that the wide dynamic range of responses to auxin may, in part, be accounted for by the range of affinities measured for different co-receptor complexes [12,14,15].

The diversity of receptors suggests some differentiation of activity as well as dose dependence [5,6,14]. It was of interest to pursue the selectivity found for picolinate herbicides and investigate the possibility of receptor sub-class-specific ligands [16]. We have explored the tetrazole functional group as a bioisostere (a chemical mimetic that sustains biologically activity) of carboxylic acids [17,18] and shown that indole-3-methyl tetrazole (compound 5, Figure 1; iMT) not only works as a weak auxin, but that it binds selectively to TIR1, and not to AFB5. Rational design was shown to improve its affinity for TIR1 without changing this selectivity.

## Results and Discussion

### Modelling alternative ligands for TIR1

Using the TIR1 crystal structure with IAA bound (PDB file 2P1P) we inspected the atomic distances from the carboxylic oxygens of IAA to nearby residues and noted distances of 3.47Å to $Arg_{403}$ and 4.52Å to $Arg_{436}$ (Figure 2A). This indicated some exploitable space in this region and so we modelled iMT (Figure 1) in the TIR1 site using the structure builder feature in Chimera [19]. Side-by-side comparison of IAA with iMT (Figure 2 A & B) show the indole rings to be superimposable, with the tetrazole group extending past the position of the carboxylic acid in IAA, further down into the pocket. In our model the atomic distances were 1.89Å to $Arg_{403}$ and 2.34Å to $Arg_{436}$. Furthermore, the proximity of the $Ser_{438}$ residue was also considered a probable hydrogen bonding partner for iMT (Figure 2B).

### Docking iMT into TIR1

Docking algorithms use more robust molecular force field calculations than a structure mutation in Chimera and so iMT and IAA were docked using the Vina algorithm [20,21] into the auxin-binding pocket of TIR1 using the coordinates from crystal structure 2P1Q (TIR1 with ligand and IAA7 degron bound, but removed for docking). We reasoned that 2P1Q would be the most appropriate template for docking in order to simulate the active state. Although there are no gross conformational changes during binding [13], there are subtle side group moves implied from the different crystal datasets. Docking predicted that the indole ring of bound iMT (scoring function -8.2 kcal/mol) would be superimposed onto that for docked (scoring function -8.1 kcal/mol) and crystallographic IAA (Figure 2C), with a slight difference between alignments of the tetrazole and carboxylic acid. Interatomic distances between the tetrazole and neighbouring arginines show 3 hydrogen bond donors within range (Figure 2D). Thus *in silico* docking predicted that iMT would bind to TIR1, making it a non-carboxylic acid auxin.

### iMT binds TIR1 *in vitro* and acts as an auxin *in vivo*

iMT was synthesised in a single step reaction from indole-3-acetonitrile with the addition of azide in a cycloaddition reaction (Supporting Information Scheme 1; physicochemical properties are also presented in Supporting Information Table 1). It was tested for binding to TIR1 using SPR. When mixed with purified TIR1, iMT supported TIR1 co-receptor assembly on the SPR chip with an activity of 18% relative to IAA (both at 50 μM; Figure 3A). We then used SPR to screen a selection of other aromatic tetrazoles, including (2-(naphthalen-1-yl)tetrazole), the tetrazole equivalent of 1-NAA (**11,**

Supporting Information Figure 1). None of this collection of compounds bound to TIR1 or AFB5, consistent with a lack of auxin activity in previous whole plant assays [22,23].

It is notable that the SPR binding signal from iMT plateaued rapidly on both association and dissociation phases (Figure 3A) unlike that for IAA, suggesting more rapid kinetics. Single cycle kinetics recorded a $K_D$ of 210 µM for iMT, compared to 5 µM for IAA (Table 1). The lower affinity for iMT is contributed by a 15-fold faster off-rate constant than for IAA, and an almost 3-fold slower on-rate constant.

The SPR assays are performed with TIR1 expressed in insect cells. In order to check activity with plant-expressed TIR1, we tested the efficacy of iMT in pull-down assays with FLAG-tagged AtTIR1 expressed in *Nicotiana benthamiana* plus IAA7 peptide [24]. We observed a weak response compared to IAA (Figure 3C), which is consistent with the SPR data on kinetic rates.

### iMT lead optimisation

Whilst iMT has been shown to be an active auxin, it is considerably weaker than IAA and many other synthetic auxins, and so rational design was applied to improve its activity. Past structure-activity relationship studies for auxins have shown that the addition of chlorine at the 4 or 6 positions of IAA yields potent auxins [25,26,27]. We also envisaged the chlorines might improve uptake properties [27]. Therefore, to start lead optimisation of iMT we synthesised the corresponding 4-, 5- and 6-monochlorinated analogues (Supporting Information Scheme 2). Binding analysis using SPR showed that 4-Cl-iMT did have enhanced binding to TIR1 (Figure 3A; Table 1. Addition of chlorine at the 6 position did not improve or reduce affinity and addition of Cl at the 5 position significantly decreased affinity for TIR1 and reduced activity *in vivo* (Table 2).

### Activity *in planta*.

Having established that iMT was active as an auxin in receptor binding assays, it was necessary to test whether or not binding translated into auxin activity *in planta*. When *DR5::GUS* Arabidopsis seedlings were grown on agar containing the test compound, iMT induced GUS activity at 50 µM, compared to IAA which gave a signal at 1 µM (Figure 4A). We followed this up with the auxin reporter *DII::VENUS* line [28] which showed a characteristic marked decrease in YFP signal in the presence of IAA. Again, iMT showed an auxin-like response, but reduced compared to that of IAA (Supporting Information Figure 2). Supporting these data, we treated 6-day-old *DR5::GFP* reporter line seedlings with IAA or iMT and recorded the GFP signal from primary root tips after 2 and 24 hours (Figure 4B). After 2 hours only IAA induced the characteristic increase in GFP signal in the epidermis, steele and the lateral root cap [29]. After 24 hours iMT-treated roots also showed an enhanced GFP signal in cells types characteristic of auxin responses, although it remained weaker than the IAA-induced signal (Figure 4B). We also noted responses to iMT analogues 4-Cl-iMT and 6-Cl-iMT in the *DR5::GUS* assays

(Figure 4A), in the *DII:VENUS* reporter assay (Supporting Information Figure 2) and in the *DR5::GFP* assay at 24 hours, especially for 4-Cl-iMT (Figure 4B). In order to reveal the timescale of responsiveness in live cells, we used a protoplast transient reporter assay [30] adapted for use with auxin-sensitive promoters and automated to record continually (Figure 4C; Supporting Information Figure 3). A series of doses of IAA or 4Cl-iMT were applied in parallel and the results indicated extended response lag times for iMT analogues compared to IAA, with responses becoming evident only after about 2 hours. This is consistent with the DR5::GFP data above (showing no response at 2 h) and might represent poor uptake kinetics for iMT and its analogues. Nevertheless, a clear, auxin-dependent signal is generated as 4Cl-iMT accumulates.

**Root growth assays quantify auxin activity**

The genetic reporter assays provided evidence that iMT was an active auxin *in vivo* and showed uptake of the compound into Arabidopsis roots and into protoplasts. However, genetic reporter assays are difficult to quantify and so we conducted dose-response assays using seedling root growth inhibition to evaluate $IC_{50}$ values for each compound (Table 2). Defining the $IC_{50}$ value as the concentration of compound needed to reduce primary root growth to 50% of the length without treatment, we obtained an $IC_{50}$ value of 46 µM (±3) for iMT compared to 41 nM (±7) for IAA in Col-0. We observed complete root growth inhibition at 300 µM for iMT compared with 11 µM for IAA (Supporting Information Figure 4). The 1000-fold difference in activity *in vivo* is greater than the difference in affinity observed for receptor binding (approximately 30-fold, Figure 3C). We noted above that the response to iMT was slower than for IAA (Figure 4C), and these extended $IC_{50}$ values also suggest that uptake of iMT by plant cells is impaired affecting potency. Nevertheless, in the same assay plates we observed increased lateral root density after iMT treatment, another characteristic auxin response (Figure 5), and so iMT is acting positively as an auxin, and not by interfering with selective elements of auxin physiology as reported for some other small molecules such as *cis*-cinnamic acid and 3,4-(methylenedioxy)cinnamic acid [31, 32].

**iMT does not bind AFB5 and this can be explained using molecular docking**

In addition to testing iMT for binding to TIR1, we investigated binding to AFB5 (Figure 3B). Somewhat surprisingly, we noted that iMT did not induce AFB5 co-receptor assembly, revealing a selectivity towards TIR1. This is the opposite selectivity to that found for picloram and the 6-arylpicolinate DAS534, which bind preferentially to AFB5 [7, 8].

In order to investigate the basis of iMT's failure to bind to AFB5 we identified the residues lining the auxin-binding pocket of each receptor [33] and aligned complementary sequences from AFB5 (Figure 6A).

There are two key changes; His78 becomes an arginine, and Ser438 becomes an alanine in AFB5. A change of histidine to arginine represents an increase in residue size and polarity, whilst a change of serine to an alanine is a decrease in polarity, an increase in hydrophobicity and loss of a key hydrogen bond acceptor. To understand how these changes might result in binding selectivity we used the crystal structure of TIR1 (PDB file 2P1Q) and the homology model built for AFB5 from this template [12] for docking *in silico* (Figure 6B to G). The His438Arg change is the most likely contributing factor to iMT receptor selectivity. IAA docked into the AFB5 pocket in a favourable pose, resembling that for TIR1, such that the aromatic ring is surrounded with hydrophobic residues parallel to the base of the pocket, and the carboxylic acid is orientated towards arginine residues at the base of the pocket, and towards the centre of the protein. Docking iMT into the AFB5 pocket suggested that the increased steric bulk of the arginine displaced the tetrazole group upwards, with this polar group now imposing on space previously taken by the alpha-carbon of IAA, and tilting the pose of the indole ring with respect to the base of the pocket. Such a pose for iMT would be likely to reduce favourable interactions with the AUX/IAA degron by perturbing the hydrophobic interactions of auxin with the WPPV motif [13]. Binding analysis of the monochlorinated iMT analogues revealed that binding remained selective for TIR1 with no binding to AFB5 (Figure 3; Table 1)

**Mutant Arabidopsis lines confirm that receptor specificity is maintained *in vivo*.**

In order to confirm that iMT is selective for TIR1 *in planta*, loss of function mutant lines of *A. thaliana* were evaluated in the root growth assay using dose-response experiments (Supporting Information Figure 4) In the *tir1-1* line we observed a shift in $IC_{50}$ from 46 µM for iMT to 90 µM (Table 2), and even at the highest dose of 300 µM primary root growth was not totally inhibited. With *afb5-5* we observed no change in the $IC_{50}$ value compared to wild type, consistent with the specificity seen for iMT and TIR1 *in vitro*. Specificity is also apparent for the lateral root growth trait (Figure 5), with the pattern of responsiveness to iMT being identical between Col-0 and *afb5-5*, but distinct from *tir1-1*. Tests were extended to quantitative RT-PCR to evaluate responsiveness using widely used auxin-responsive genes (Figure 4D). Treatment with IAA (1 µM) and 4-Cl-iMT (10 µM) gave similar auxin-like responses in Col-0, but the response to 4-Cl-iMT was absent in *tir1-1* for all three reporter genes, consistent with binding selectivity of 4-Cl-iMT for TIR1. The reporters *GH3.3* and *GH3.1* gave partial responses to 4-Cl-iMT in *afb5-5*, whilst reporter *IAA5* responded as for IAA, as anticipated for a line with a functional TIR1. Selectivity for TIR1 was also shown to be maintained with lateral root density in the *tir1-1* line being significantly reduced compared to Col-0 and *afb5-5* at the active higher dose rates (Figure 5,Table 2). Picloram demonstrated its inverse selectivity and preference for AFB5 (Figure 5A).

**Discussion**

We have demonstrated that isosteric replacement of the carboxylic acid on IAA with a tetrazole yields an active auxin that binds to the auxin receptor TIR1. Further, this isosteric change confers on iMT selectivity for TIR1 which introduces the first auxin that does not engage with the receptor homologue AFB5. Isosteric replacements have been reported for auxin previously, including the report of weak activities in the *Avena* coleoptile extension and pea split epicotyl assays [22, 23], but never pursued. The tetrazoles of 1-naphthylacetic acid and 2,4-dichlorophenoxyacetic acid were found to be far weaker than iMT, as were other bioisosteres evaluated at the same time. Given the weak activity, perhaps it is not surprising that no further interest has been shown in the 60 years since. However, with an SPR binding assay, as well as Arabidopsis *TIR1/AFB* mutants, we were able to revisit the activity of non-carboxylic synthetic auxins and this revealed target site specificity more extreme than the known prefence of AFB5 for picloram and the picolinate auxins [7, 8]. This is the first report of a chemical tool for examining receptor-selective responses and redundancy within the TIR1 family of receptors.

We have suggested the molecular basis of iMT specificity to be a histidine the base of the TIR1 binding pocket which is replaced by an arginine residue in AFB5 which protrudes more (Figure 6). Without the crystal structure of the AFB5 receptor this explanation is based on a homology model [12], although this remains a reasonable hypothesis until the AFB5 structure is solved. We recognise that we have not represented the AFB2 clade in this study although it clusters close to TIR1 in sequence alignments [4]. It is clear from all the data collected from *tir1-1* plants that the effects of iMT and 4-Cl-iMT are dominated by recognition through TIR1. Nevertheless, we can't exclude some efficacy with AFB2 and we note that at the relatively high concentrations needed for activity *in vivo* with this current generation of iMT analogues there could be some cross-over signalling from other members of the TIR1 family. Importantly, we now have chemical templates suitable for TIR1-dominant (iMT) and AFB5-dominant (picolinate) activation of auxin signals. We do not yet fully understand the significance of six redundant auxin receptors, but these compound families offer new tools to help differentiate the receptors.

The tetrazole bioisostere iMT is not as potent as IAA, with a 40-fold poorer affinity for receptor binding (Table 1). The difference in activity *in planta* is greater, perhaps due in part to reduced uptake and transport. Some reduced uptake capacity is suggested in the timecourse of the protoplast gene reporter assays (Figure 4C, Supporting Information Figure 2). Our initial lead optimisation programme has yielded increased potency for 4-Cl-iMT with an $IC_{50}$ value of 19 μM in the primary root growth inhibition assay (Table 2). In terms of utility, we may compare this to picloram which is a successful commercial herbicide. Picloram has an $IC_{50}$ of 5 μM (Table 2). A further comparison may be drawn to the herbicide glyphosate, which has an $IC_{50}$ of 11 mM on its target enzyme 5-enolpyruvylshikimate-3- phosphate synthase [34] and an $IC_{50}$ on Arabidopsis rosettes of 50 μM, with plants able to grow through treatments at 5 mM [35].

As with all agrochemicals, concerns are growing over resistance to current actives. The auxin family of herbicides faces the same challenges. Resistance has not become a global threat [36, 37] but applications are rising with the advent of dicamba- and 2,4-D-tolerant GM crops [38] [39]. However, as well as maintaining the utility of the current arsenal, binding site variation might open the door to new herbicide selectivities, wider or more restricted than the known broad-leaved dynamic of most current compounds.

**Methods**

*In silico* modelling, chemical and protein visualisation

*In silico* modelling, molecular graphics and analyses were performed with the open source UCSF Chimera package developed by the Resource for Biocomputing, Visualization, and Informatics at the University of California, San Francisco (supported by NIGMS P41-GM103311). [19]. Marvin was used for drawing, displaying and characterizing chemical structures, substructures and reactions. Calculator Plugins were used for structure property prediction and calculation (Marvin v15.10.12.0, 2015; ChemAxon (http://www.chemaxon.com). Chemical structures were drawn using ChemDraw Professional v15.0.0.106. Docking was performed using an automated docking script [21] based on the Vina algorithm [20]. Crystal structures 2P1P and 2P1Q [13] were sourced from RSCB [40].

Recombinant expression:

Expression constructs for both TIR1 and AFB5 were engineered to give fusion proteins His10-MBP-(TEV)-FLAG-TIR1 and His-MBP-(TEV)-FLAG-AFB5 respectively and were coexpressed with His10-(TEV)-ASK1 as descrbed in [8]. Generation of recombinant virus, quantification, selection, expression screening, and generation of high-titer viral stock was done by Oxford Expression Technologies (Oxford, U.K.). *Trichoplusia ni* (*T. ni High 5*) was used throughout as the host cell line for expression. Cell densities were determined with a haemocytometer (Marienfeld, Neubauer-improved 0.1mm, catalogue number: 0640030) using a 10x objective lens under a light microscope. Cells were infected at a density of 1 x$10^6$ cells/mL with multiplicity of infection of 5. The cells were harvested by centrifugation 72 h after infection and stored at -80ºC.

Cell Lysis and protein extraction:

Frozen TIR1/ASK1 and AFB5/ASK1 pellets were thawed at room temperature and lysed for 40 minutes whist rolling at 4ºC in Cytobuster™ Lysis medium (Invitrogen 5 mL per 250 mL of cell lysate) supplemented with DNAse I (Roche), protease inhibitors (cOmplete™ Protease Inhibitor Cocktail Tablets, Roche), 50 µM phytic acid (Sigma) and 1 mM reducing agent TCEP (Tris(2-carboxyethyl)phosphine hydrochloride – Sigma). The lysate was diluted upto 30mL into Buffer A (20 mM Tris-HCl pH 7.4, 200 mM NaCl, 1mM EDTA, 50 µM phytic acid, 1 mM TCEP) and was subjected to 3 x 30 seconds ultrasonication before centrifugation at 20,000 rpm at 4ºC for 15 minutes. The supernatant was then systematically filtered through 0.45 µm and 0.2 µm Whatman GD/X syringe filters.

Protein purification:

The filtered lysate was loaded onto a nickel immobilised metal affinity chromatography column (cOmplete His-Tag Purification Resin – Roche), washed with 10 column volumes of Buffer-A before elution with Buffer-B (20 mM Tris-HCl pH 7.4, 200 mM NaCl, 1mM EDTA, 50 μM Phytic acid, 1 mM TCEP, 250 mM Imidazole). TEV protease was added to the elute and incubated with mixing at 4°C overnight. The solution was then loaded onto an anti-FLAG-affinity resin (ANTI-FLAG® M2 Affinity Gel - Sigma), washed with 10 column volumes of Buffer-C (10 mM HEPES pH7.4, 150 mM NaCl, 3 mM EDTA, 50 uM Phytic acid, 1 mM TCEP, 0.05% Tween 20) and eluted with 3X-FLAG peptide (Sigma) at 100ug/mL. Protein was stored on ice and protein concentrations were assayed by nanodrop $A_{280}$ nm measurement (Thermo Scientific).

SPR assays:

The auxin binding assays using SPR were done on a Biacore 2000 instrument as described previously [8,41]. The kinetic analysis of iMT analogues was perfomed by single cycle kinetics on a Biacore T200, titrating the compounds mixed with constant TIR1 before injection onto the SA chip. The orientation of the assay was otherwise as for the Biacore 2000 assays, and degron peptide density on the chips was controlled so that Rmax > 300 RU.

Pull-down assay:

The pull down assay was done according to the method described in [24] where *Nicotiana benthamiana* leaves were infiltrated with agrobacterium to express FLAG-TIR1 and leaf lysates were incubated with biotynalted AUX/IAA7 degron (biotinyl-AKAQVVGWP PVRNYRKN) attached to streptavidin beads. The reactions were done in the absence and presence of IAA or iMT at specified concentrations.

Plant assays:

All root growth assays were done in Col-0 and mutants in this background, *tir1-1* and *afb5-5* lines [2,27]. A series of plates with 15 serial dilutions for each compound was prepared in half strength Murashige and Skoog medium. From the top concentration of 300 μM a three-fold dilution series was prepared, plus a control without compound, giving a total of 16 plates.

From a pool of 6 day old seedlings a random selection of 10 were transferred onto each of the treated plates and the position of the primary root tip was marked. The plates were placed randomly in the stack such that the concentrations were not in order to account for positional bias. A plate with no treatment was included in every assay for comparison with the lower end of the dose response (longest roots) and as an indicator of reproducibility in the assay. The stack was placed with the seedlings vertical for 6 days at 12 hour day (22 °C) 12 hour night (18 °C) cycles then scanned (HP PSC 2500 series) at 1200 dpi in colour mode. Primary root growth from the marked point was measured in Image J [42] and plotted using GraphPad Prism v 7.0.

The primary root growth measurements were fitted to a logistic function (2.1), using the Levenberg–Marquardt algorithm, in QTI plot (IONDEV SRL, Romania, v 0.9.8.9). The standard deviation of the data points was weighted into the algorithm $f(x) = \frac{M}{1+e^{-k(x-x_0)}}$ where: M = Maximal value on curve, e = natural logarithm, $x_0$ = $IC_{50}$ value, k = Steepness of the curve.

Lateral root hairs were counted on screen and statistical comparisons were conducted in GraphPad Prism v 7.0 using a two-way ANOVA looking at comparisons within each row (i.e. compound concentration µM) and comparing columns (i.e. mutant lines) to a reference control column (i.e. WT line), multiple comparisons reported to a 95% confidence level (Supporting Information Table 2).

*DR5::GUS* reporter assays:

5-days-old *DR5::GUS* seedlings were cultured in liquid MS medium (1.2% sucrose) with chemicals for 16h at 23ºC. The seedlings were then washed with a GUS staining buffer [43] and transferred to a GUS staining buffer containing 1 mM X-gluc. The seedlings were then incubated at 37 °C until sufficient staining developed.

*DII::VENUS* assay:

5-day-old *DII::VENUS* seedlings [28] were cultured in liquid MS medium (1.2% sucrose) with 20 μM yucasin, IAA biosynthesis inhibitor for 6h at 23ºC to accumulate DII-VENUS protein. The chemicals were then added into culture medium at indicated concentration. After 1h incubation at 23 ºC in dark, DII-VENUS image was captured by fluorescent microscopy BX-50 (Olympus, Japan) with YFP filter sets.

*DR5::GFP* reporter assay:

Col-0 *DR5::GFP* seedlings [44] were germinated as above for root growth assays and used directly from the plates. Seedlings from ther same batch of germinants were placed onto fresh media (half-strength MS) prepared with compound from 100mM stocks in DMSO to give a final concentration of 50μM (final DMSO concentration of 0.05% v/v) in 6-well plates for 2 and 24 h in the dark. Treatments were started simultaneously. At sampling, primary roots were cut to 3 cm, treated with 10 ug/ml Propidium Iodide, then placed onto a slide in water and imaged using a Leica (Germany) LSM 880 imaging system controlled by Leica Zen software with a 25 x oil immersion objective. GFP was excited with a 488 nm laser line and detected between 499 nm and 544 nm. PI was excited with a 514 nm laser and detected between 598 nm and 720 nm. Control roots (DMSO 0.05% v/v) were used to benchmark imaging settings and the same imaging parameters were used for both days, except for the IAA treatment at 24 hours for which the gain was lowered to obtain a non-saturated image.

Protoplast reporter assay:

Mesophyll protoplast were obtained from leaf 8 of 4-week old plants following the "tape sandwich" method using 4 plants per genotype [45]. The IAA5 promoter (At1g15580, 920 bp) was amplified from genomic Col-0 DNA (primers 5'CCTGCAGGCTCTAGAGGATCCGCTGTCCATTATCACAAAGTC3' and 5'TGTTTTTGGCGTCTTCCATGGCTTTGATGTTTTTGATTGAAAG3'). The backbone for the pIAA5::LUC construct was generated from pFRK1::LUC (ABRC CD3-919) by digestion with BamHI and NcoI and gel-purified. Backbone and PCR-amplified IAA5 promoter were combined in a Gibson reaction using the CloneEZ kit (GenScript, USA). Plasmids containing the construct *pGH3.3::LUC* [46] or *pIAA5::LUC* were transfected together with reporter *pUBQ10::GUS* for normalisation [47]. After overnight incubation, protoplasts were treated with IAA or 4-Cl-iMT at the concentrations described and LUC substrate luciferin added. The plate was immediately placed in a Photek dark box and imaged with a photon-sensitive camera HRPCS218 (Photek; Supporting Information Figure 4) for 6 h using the software Image32 (Photek) to integrate photon capture. After imaging protoplasts were lysed for GUS activity analysis [46]. Images were processed with the Image32 software binning the photons captured for each minute into a time resolved image (TRI). Then, for each well total intensity values were extracted. Light intensity was normalised to GUS activity.

Quantitative RT-PCR:

Leaf number 8 from 4-week old plants was syringe-infiltrated with either 1% DMSO in water (mock), 10 µM 4-Cl-iMT or 1 µM IAA for 3 hours. RNAs were extracted with RNeasy Plant Mini Kit (Qiagen) and treated with TURBO™ DNase (Ambion) following manufacturer's instructions. For cDNA synthesis 1 µg of RNA was reverse-trancribed with the SuperScript™ II Reverse Transcriptase (Thermo Fisher Scientific), following manufacturer's specifications, using a primer for polyA tails d(T)$_{19}$. qPCR was performed with SYBR® Green JumpStart™ Taq ReadyMixTM (Sigma), following the manufacturer's recommendations (qPCR primers see Supporting information Table 3). All qPCR primers were tested for efficiency on a standard curve. Three technical replicates were used for each sample and 384-well plates were read using a CFX384 Touch™ Real-Time PCR Detection System (Bio-Rad Laboratories). qPCR $C_T$ values were exported into an excel file and analyzed using the $\Delta\Delta C_T$ method [48]. Data was normalised to *UBC* (AT5G25760; [49]). *UBC* expression was found to be stable under the conditions studied (Supporting information Figure 5).

Chemical synthesis:

<u>iMTsynthesis</u> via 1,3-dipolar cycloaddition [50] of sodium azide (NaN$_3$) and 2-(1H-indol-3-yl)acetonitrile with NaN$_3$, AlCl$_3$.THF in THF for 18h at 71˚C (Compound (1), Supporting Information Scheme 1 [51] [52]. 4-chloroindole-3-acetonitrile, 5-chloroindole-3-acetonitrile and 6-chloroindole-3-acetonitrile were synthesized according to the published procedure in [53], their corresponding 4/5/6 CL-iMT analougues were synthesised with the NH$_4$Cl, NaN$_3$ into DMF at 120˚C for 30h (Supporting Information Scheme 2). Complete methodologies including NMR and MS data are presented in the Supporting Information (Supporting Information Schemes 1 and 2).

**Acknowledgements:** This work was funded by the Biotechnological and Biological Research Council BBSRC of the United Kingdom (MIBTP award to MQ, BB/L009366 to RN and SK). We thank M. Estelle for providing the *afb5-5* knockout line used in this work. We acknowledge R. Schäfer for invaluable advice in troubleshooting root growth assays and confocal imaging. The protoplast assays were developed under award BB/M017982/1 from BBSRC and EPSRC for the Warwick Integrative Synthetic Biology Centre.

List of Supporting Information

Scheme 1: Synthesis of iMT

Scheme 2: Synthesis of monochlorinated analogues of iMT

Table 1: Physiochemical properties of IAA and iMT analogues

Figure 1: Additional tetrazoles tested for auxin-like activity

Figure 2: Imaging of auxin responses using genetic reporter DII::VENUS

Figure 3: Protoplast activity assays

Figure 4: Root growth inhibition curves

Figure 5: *UBC* expression observation

Table 2: Two-way ANOVA for lateral root densities under 4-CL-iMT

Table 3: qPCR reference genes and primer sequences

This material is available free of charge via the internet at http://pubs.acs.org

END OF TEXT

Tables:

| Compound | TIR1 | | | | | AFB5 | | | | |
|---|---|---|---|---|---|---|---|---|---|---|
| | ka (1/Ms) | ± (fit error) | kd (1/s) | ± (fit error) | KD (µM) | ka (1/Ms) | ± (fit error) | kd (1/s) | ± (fit error) | KD (µM) |
| IAA | 8.06E+02 | 3.4 | 3.85E-03 | 9.60E-06 | 4.78 | 4.61E+02 | 6 | 7.02E-02 | 4.50E-04 | 154 |
| iMT | 3.04E+02 | 3.2 | 6.39E-02 | 1.40E-04 | 210 | - | - | - | - | - |
| 4-Cl-iMT | 2.43E+02 | 1.8 | 3.82E-02 | 1.90E-04 | 157 | - | - | - | - | - |
| 5-Cl-iMT | 3.28E+02 | 8.8 | 1.20E-01 | 5.30E-04 | 366 | - | - | - | - | - |
| 6-Cl-iMT | 3.75E+02 | 4.7 | 7.64E-02 | 2.20E-04 | 204 | - | - | - | - | - |

**Table 1. Kinetic binding data.**
The Biacore T200 single cycle kinetic facility was used with titrations of each compound against constant receptor concentration to derive kinetic values for each compound.

| | A. thaliana lines (IC$_{50}$ values in μM) | | | | | | |
|---|---|---|---|---|---|---|---|
| Compound | Col-0 | ± SE | *tir1-1* | ± SE | *afb5-5* | ± SE | Sensitivity |
| IAA | 0.04 | -0.01 | 0.06 | -0.01 | 0.03 | -0.01 | - |
| 4-Cl-IAA | 0.04 | -0.004 | - | - | - | - | - |
| 5-Cl-IAA | 0.19 | -0.01 | - | - | - | - | - |
| 6-Cl-IAA | 0.02 | -0.002 | - | - | - | - | - |
| Picloram | 4.79 | -0.87 | 8.4 | -0.84 | 43.66 | -6.37 | AFB5 |
| iMT | 46.21 | -2.87 | 90.61 | -6.99 | 45.59 | -4.9 | TIR1 |
| 4 Cl iMT | 19.05 | -1.96 | 36.91 | -3.92 | 19.12 | -3.39 | TIR1 |
| 5 Cl iMT | 59.47 | -6.85 | 55.04 | -4.29 | - | - | - |
| 6 Cl iMT | 32.48 | -1.25 | 34.82 | -2.17 | 28.09 | -1.71 | TIR1 |

**Table 2: iMT analogues are active as auxins in the Arabidopsis primary root growth inhibition assay.**

The IC$_{50}$ values (±SE) for the inhibition of primary root growth were calculated from statistical fits to dose response data. Receptor preferences are noted on the right.

Figures:

**A**

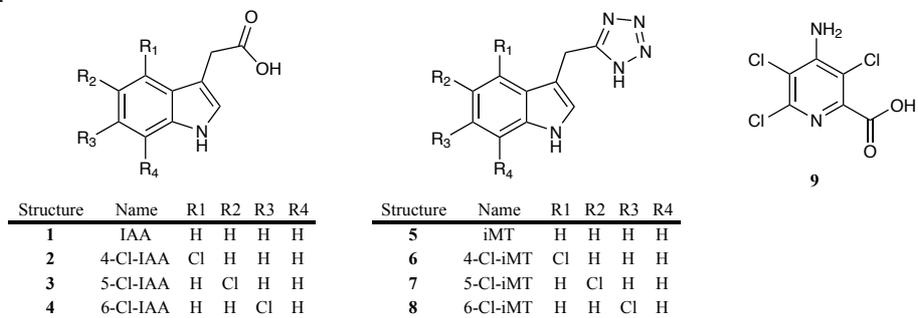

| Structure | Name | R1 | R2 | R3 | R4 |
|---|---|---|---|---|---|
| 1 | IAA | H | H | H | H |
| 2 | 4-Cl-IAA | Cl | H | H | H |
| 3 | 5-Cl-IAA | H | Cl | H | H |
| 4 | 6-Cl-IAA | H | H | Cl | H |

| Structure | Name | R1 | R2 | R3 | R4 |
|---|---|---|---|---|---|
| 5 | iMT | H | H | H | H |
| 6 | 4-Cl-iMT | Cl | H | H | H |
| 7 | 5-Cl-iMT | H | Cl | H | H |
| 8 | 6-Cl-iMT | H | H | Cl | H |

**Figur**
Chem
analo

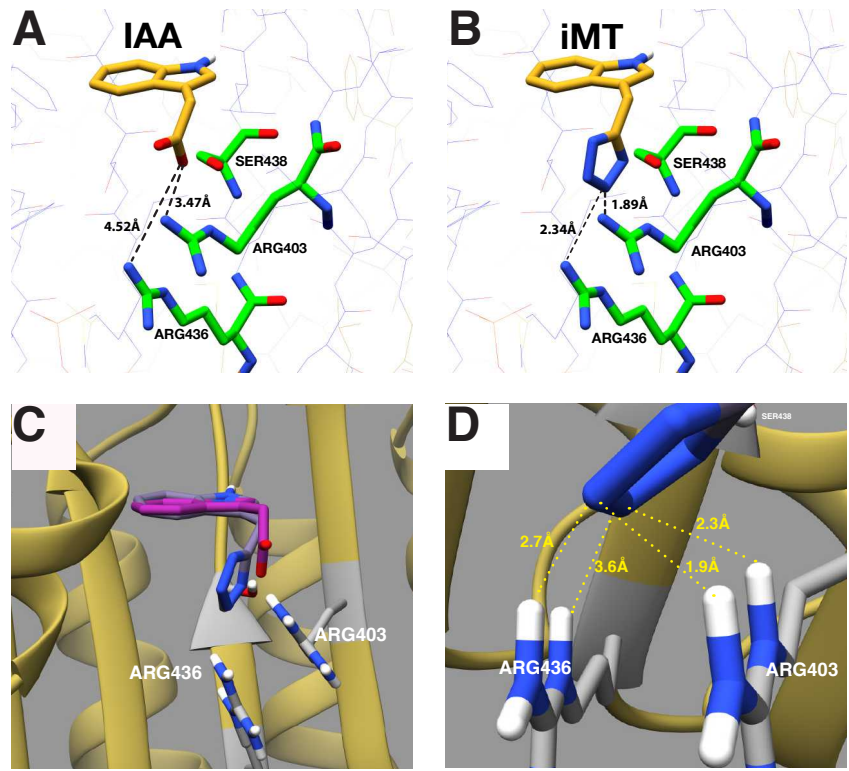

**Figure 2. Modelling iMT as a ligand for TIR1.**
A: The crystal structure of IAA (gold) bound in the TIR1 receptor, with key residues shown with coloured heteroatoms. Interatomic distances are shown as dashed black lines. B: As A, but with the tetrazole analogue modelled in the binding site using Chimera. C: IAA (magenta) and iMT (grey), each docked (AutoDock Vina) in the deep binding pocket of TIR1 (gold, ribbon). The docked poses are consistent with the modelling in B. The indole ring of iMT adopts the same plane as that for IAA (which is the same as that seen in the crystal structure), with the tetrazole group projecting past the carboxylic acid group of IAA. D: shows the atomic distances between the tetrazole group nitrogens and neighbouring arginine residues.

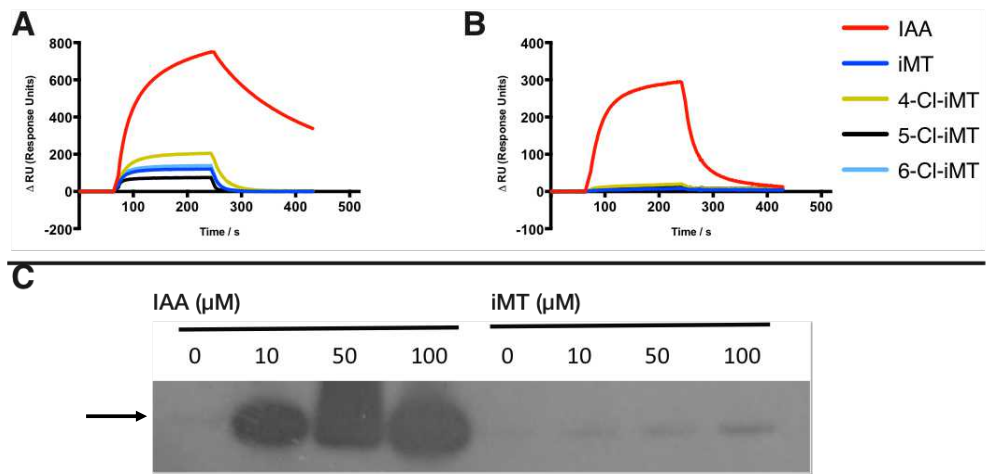

**Figure 3. iMT binds to TIR1, but not to AFB5.**
SPR binding curves for compounds screened at 50μM against TIR1 (Panel A) and AFB5 (Panel B). In each case IAA (red trace) is used for reference. With TIR1 we observe a saturating binding response with a rapid off rate for all iMT analogues. None of the iMT analogues were active against AFB5. C: A pull-down assay for FLAG-TIR1 in the presence of increasing concentrations of IAA (left) and iMT (right). A western blot developed with anti-FLAG antibody detects FLAG-TIR1 (arrow) bound to streptavidin-coated beads loaded with biotinylated degron peptide. As with SPR, there is a strong response with IAA, and the iMT response is dose-dependent, but weaker.

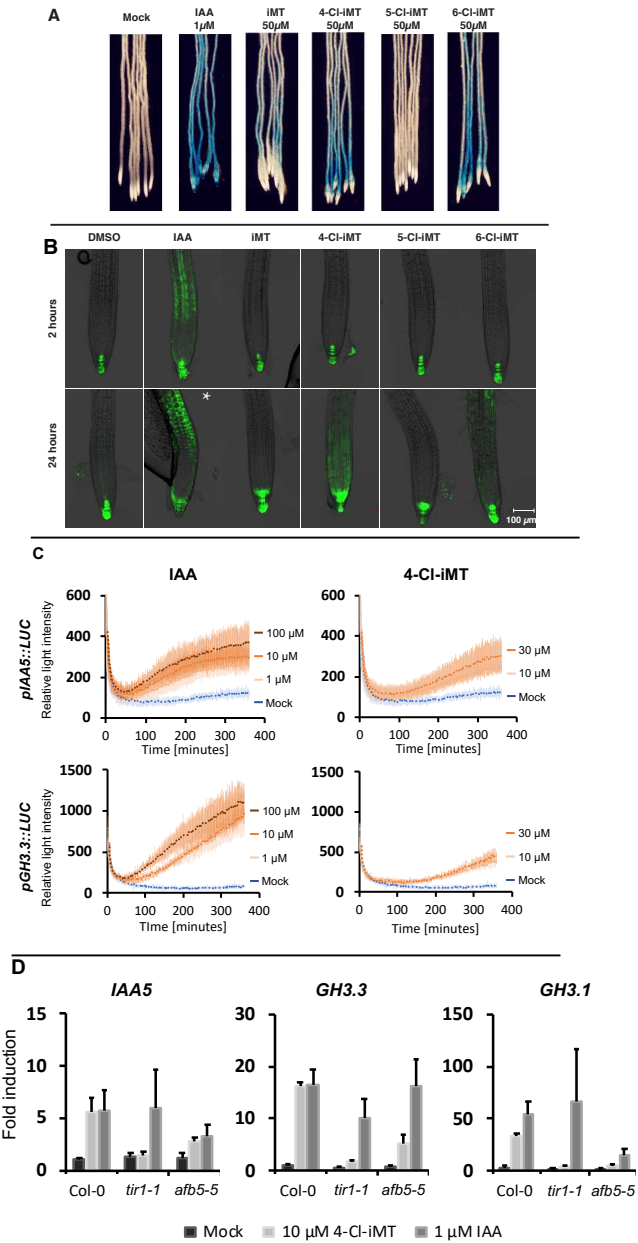

**Figure 4. iMT analogues are active auxins in vivo.**
A: The DR5::GUS reporter line indicates auxin activity of iMT and its monochlorinated analogues. Treatments were for 16 h at 1μM for IAA, 50 μM for iMT and its analogues. No activity was seen with 5-Cl-iMT. B: Signals from DR5::GFP after 2 and 24 hours of treatment. Activity is seen after 2 hours with IAA, iMT and 4- and 6-Cl-iMT show activity within 24 hours, with 4-Cl-iMT giving a comparatively strong response. *For the image of IAA treatment after 24 hours the gain was lowered to avoid a saturated signal. C: Arabidopsis protoplasts were transformed with auxin-sensitive reporter constructs pGH3.3::LUC (above) or pIAA5::LUC (below) before treatment with IAA (left) or 4-Cl-iMT (right), each over responsive dose ranges and each with a mock treatment (blue). Luciferase activity was recorded each minute for 6 hours. Error bars represent standard deviations of technical replicates and the plots are a representative set from three biological repeats. D: Quantitative PCR data for three auxin-responsive reporter genes, *IAA5*, *GH3.3* and *GH3.1*. RNA samples were prepared from treated leaf tissues collected 3 hours after treatment with mock, 4-Cl-iMT or IAA (10 and 1 μM respectively) from Col-0, *tir1-1* and *afb5-5* lines. Error bars represent the standard error of the mean of biological repeats (n = 4).

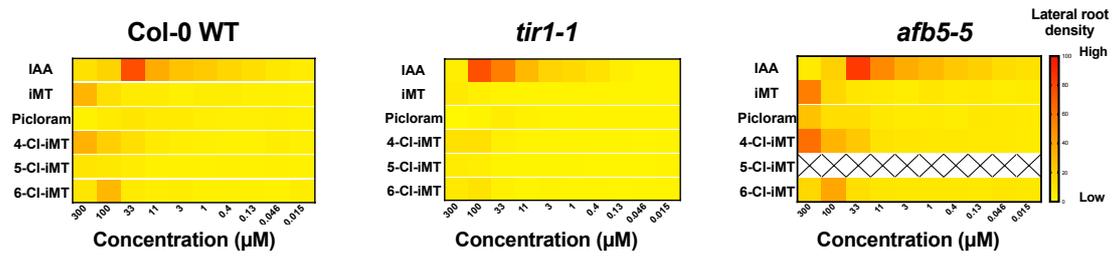

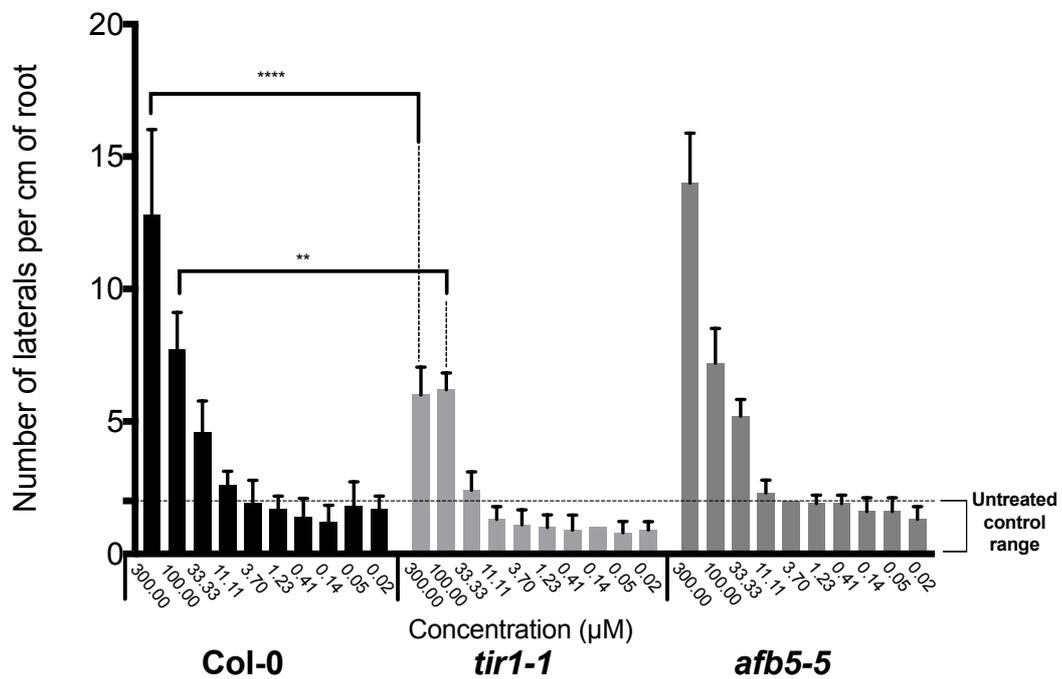

**Figure 5: The *tir1-1* mutant is insensitive to iMT and 4-Cl-iMT**

A: Normalised Lateral root densities plotted as heat maps to display the effects of compound and compound concentration for wild type, *tir1-1* and *afb5-5* knockout lines. Lateral root density is reduced in the *tir1-1* mutant challenged with iMT, 4-Cl-iMT and 6-Cl-iMT, but not in *afb5-5* nor Col-0. Cross-hatching (XX) indicates that this was not tested. B: From the same data as A but focussing on lateral root densities for lines treated with 4-Cl-iMT. Error bars indicate +/- the standard error of the means. A two-way ANOVA was used to compare densities at each concentration for each mutant line versus Col-0; ** P≤0.01 and **** is for P≤0.0001 using 10 replicates per concentration per seed line. The loss of density at higher concentrations is seen only with *tir1-1* and not in the *afb5-5* line.

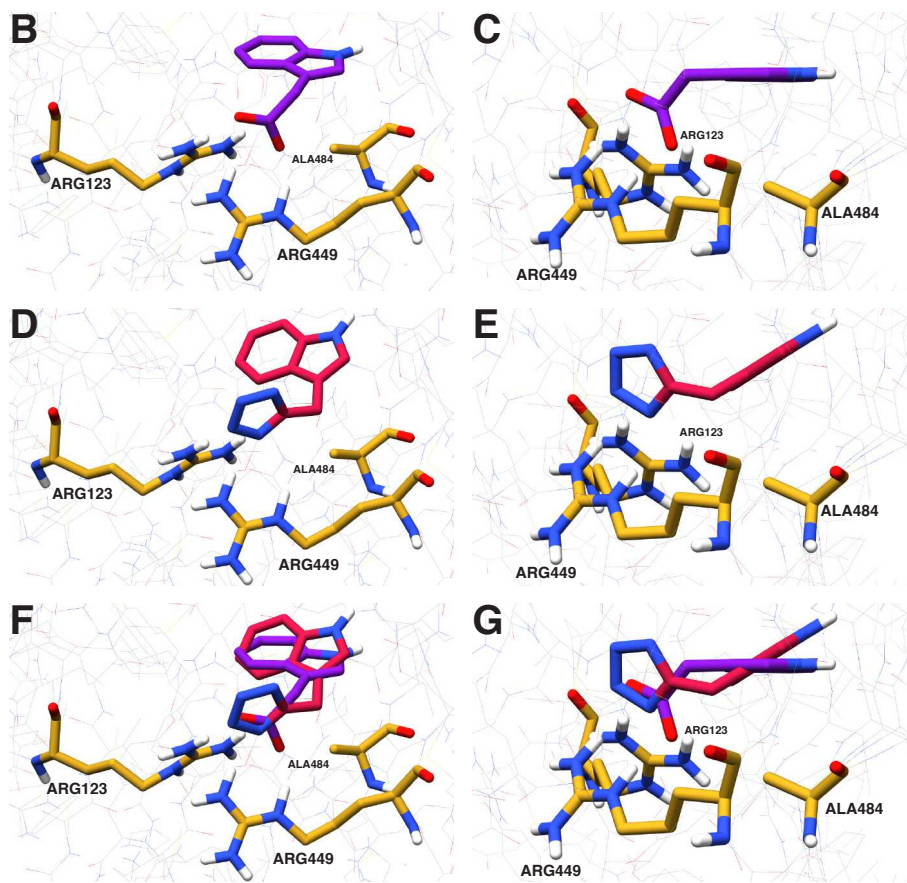

**Figure 6. A model for iMT selectivity based on space constraints in AFB5.**
**A:** The binding pocket residues of TIR1 are alligned against those of AFB5, using the residue numbers for TIR1 (Clustal 2.1). Key differences are highlighted in yellow and include His78, which is Arg in AFB5, and S438 Ala in AFB5. **B – G: Views of the homology model for AFB5** (Calderon-Villalobos et al., 2012) showing IAA (B and C) or iMT (D and E), or both F and G) docked using AutoDock Vina. The views in each left hand panel are similar to those for TIR1 in Figure 1, and on the right views are revolved to show the pose of the side group out of the plane of the aromatic ring system**.** Note that iMT does not adopt the same pose as IAA in AFB5 because the space occupied by the tetrazole in TIR1 (Figure 1) is partially occupied by Arg123 in AFB5, forcing the tetrazole up and away from the base of the pocket, resulting in the indole twisting from alignment with the base of the pocket.

# SUPPORTING INFORMATION

# The tetrazole analogue of the auxin indole-3-acetic acid binds preferentially to TIR1 and not AFB5.

Mussa Quareshy [1†], Justyna Prusinska [1], Martin Kieffer [2], Kosuke Fukui [3], Alonso J. Pardal [1], Silke Lehmann [1,5], Patrick Schafer [1,5], Charo I. del Genio [1], Stefan Kepinski [2], Kenichiro Hayashi [3], Andrew Marsh [4], Richard M. Napier [1†]

[1] School of Life Sciences, University of Warwick, Coventry, CV4 7AL, UK

[2] Centre for Plant Sciences, University of Leeds, Leeds LS2 9JT

[3] Department of Biochemistry, Okayama University of Science, 1-1 Ridaicho, Kita-ku, Okayama-shi Okayama, JP 700-0005

[4] Department of Chemistry, University of Warwick, Coventry, CV4 7AL, UK

[5] Warwick Integrative Synthetic Biology Centre

† Corresponding authors:

mussaquareshy@gmail.com; Richard.napier@warwick.ac.uk

List of Supporting Information

Scheme 1: Synthesis of iMT

Scheme 2: Synthesis of monochlorinated analogues of iMT

Table 1: Physiochemical properties of IAA and iMT analogues

Figure 1: Additional tetrazoles tested for auxin-like activity

Figure 2: Imaging of auxin responses using genetic reporter DII::VENUS

Figure 3: Protoplast activity assays

Figure 4: Root growth inhibition curves

Figure 5: *UBC* expression observation

Table 2: Two-way ANOVA for lateral root densities under 4-CL-iMT

Table 3: qPCR reference genes and primer sequences

# Supporting Information Scheme 1

Synthesis of iMT, 3-((1H-tetrazol-5-yl)methyl)-1H-indole

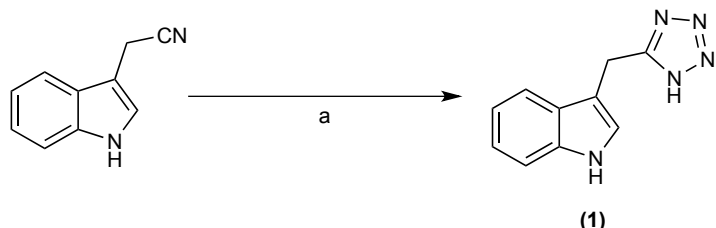

(1)

Scheme 1. *Reagents and conditions*: (a) NaN$_3$, AlCl$_3$.THF, THF, 71°C, 18h.

Based on methodology from (McManus & Herbst, 1959; Dolusić et al., 2011), to a stirred solution of 2.6mL 0.5M AlCl$_3$.THF in a 2-neck round bottom flask at 0°C in an ice bath was added NaN$_3$ 180mg (3.5 mmol), left for 2 hours and kept under an inert N$_2$ atmosphere. Indole-3-acetonitrile powder 190 mg (1.25 mmol) was dissolved in 5 mL THF and added via the side neck, and the residual powder in the neck was washed down with THF 2 mL. The mixture was taken off ice and placed onto a stirrer hotplate and allowed to reflux for 18 hours with a water condenser column attached. Teflon tape was used to seal the glassware connections. After 18 hours, the mixture was left to cool where we observed an off-white creamy precipitate. The aluminate precipitate was dissolved into 20 mL of 1M citric acid, washed with 3 x 25mL of ethyl acetate (EtOAc), then 1 x 25 mL water, followed by 1 x 25 mL brine and 1 x 25 mL water. The extract was then dried over MgSO$_4$. The excess solvent was evaporated off in a rotary evaporator, producing residual pale yellow oil. The crude material was dissolved in 5 mL EtOAc and purified by flash chromatography on standard 60Å (Sigma Aldrich) silica gel with an EtOAc:MeOH (95:5) gradient set up. Fractions containing the product were pooled, concentrated with a rotary evaporator and left to dry in a vacuum oven for 24 hours giving an off white powder (60 mg, 24%)

1H NMR (300MHz ,DMSO-d$^6$) δ = 15.92 (s, 1 H), 11.00 (s, 1 H), 7.41 (d, J = 8.1 Hz, 1 H), 7.36 (d, J = 8.0 Hz, 1 H), 7.25 (s, 1 H), 7.08 (t, J = 8.1 Hz, 1 H), 6.97 (t, J = 7.49 Hz, 1 H), 4.36 (s, 2 H) $^{13}$C NMR (75 MHz, DMSO-d6) δ = ppm 161.3, 135.7, 125.9, 123.3, 120.7, 118.1, 117.5, 111.0, 107.7, 18.9 HRMS (m/z): [M]$^+$ calcd. For C$_{10}$H$_{10}$N$_5$, 200.0931; found, 200.0932

# Supporting Information Scheme 2

The synthesis of monochloroniated indole-3-methyltetrazole analogues

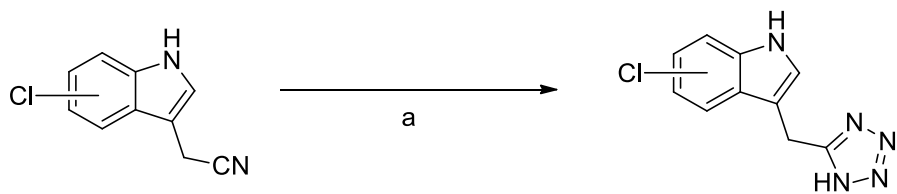

Scheme 2. *Reagents and conditions*: (a) NH₄Cl, NaN₃, DMF, 120°C, 30h

4-Chloroindole-3-acetonitrile, 5-Chloroindole-3-acetonitrile and 6-Chloroindole-3-acetonitrile were synthesized according to the published procedure in (Katayama, 2000; Luo et al., 2014)

(6) (4-Cl-iMT) 3-((1*H*-tetrazol-5-yl)methyl)-4-chloro-1*H*-indole

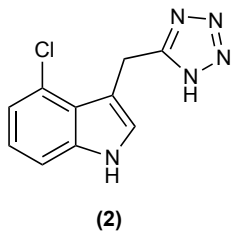

(2)

To the solution of 4-chloroindole-3-acetonitrile (120 mg, 0.63 mmol) in dimethylformamide (3mL) was added ammonium chloride (133 mg, 2.50 mmol) and sodium azide (100 mg, 1.56mmol). The reaction mixture was then stirred for 30 h at 120°C. The resulting solution was cooled and added to water (30 mL), and extracted with ethyl acetate [EtOAc] (20 mL × 2). The organic layer was washed with saturated NH₄Cl solution and brine, and then dried over Na₂SO₄. The residue was purified by a silica gel column chromatography (Chloroform:acetone=4:1) to give 3-((1H-tetrazol-5-yl)methyl)-4-chloro-1H-indole (1) as white powder (62 mg, 42% yield): 3-((1H-tetrazol-5-yl)methyl)-4-chloro-1H-indole (4-Cl-iMT)

$^1$H NMR (500 MHz, DMSO-d₆) δ = 11.40 (s, 1H), 7.37 (d, J = 8.9 Hz, 1H), 7.35 (s, 1H), 7.06 (t, J = 7.7 Hz, 1H), 6.97 (d, J = 7.5 Hz, 1H), 4.55 (s, 2H), $^{13}$C NMR (126 MHz, DMSO-d6) δ = 156.3, 138.0, 126.6, 124.3, 123.2, 122.1, 119.4, 111.0, 107.7, 20.9; HRMS (*m/z*): [M-H]⁻ calcd. For C₁₀H₇ClN₅, 232.0395; found, 232.0393. FAB-MS (*m/z*): [M+H]⁺ 234.

(7) (5-Cl-iMT) 3-((1*H*-tetrazol-5-yl)methyl)-5-chloro-1*H*-indole

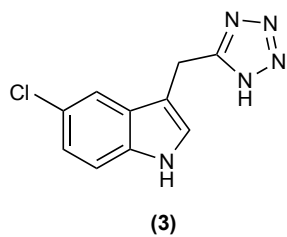

(3)

To the solution of 5-chloroindole-3-acetonitrile (250 mg, 1.32 mmol) in dimethylformamide (5mL) was added ammonium chloride (280 mg, 5.31 mmol) and sodium azide (212mg, 3.23 mmol). The reaction mixture was then stirred for 25 h at 120˚C. The resulting solution was cooled and added to water (40 mL), and extracted with ethyl acetate [EtOAc] (30 mL × 2). The organic layer was washed with saturated NH$_4$Cl solution and brine, and then dried over Na$_2$SO$_4$. The residue was purified by a silica gel column chromatography (Chloroform:acetone=4:1) to give 3-((1H-tetrazol-5-yl)methyl)-5-chloro-1H-indole (**2**) as white powder (116 mg, 38% yield):

$^1$H NMR (400 MHz, acetone-d$_6$) δ = 10.43 (s, 1H), 7.55 (d, J = 2.3 Hz, 1H), 7.43(d, J = 6.4 Hz 1H), 7.42 (s,1H), 7.10 (dd, J = 6.4 and 2.3 Hz, 1H), 4.48 (s, 2H); $^{13}$C NMR (100 MHz, acetone-d$_6$) δ = 155.9, 135.3, 128.2, 125.8, 124.5, 121.8, 117.8, 113.0, 108.8, 19.6; HRMS (*m/z*): [M]$^-$ calcd. For C$_{10}$H$_7$ClN$_5$, 232.0395; found, 232.0396.
FAB-MS (*m/z*): [M+H]$^+$ 234.

(8) (6-Cl-iMT) 3-((1*H*-tetrazol-5-yl)methyl)-6-chloro-1*H*-indole

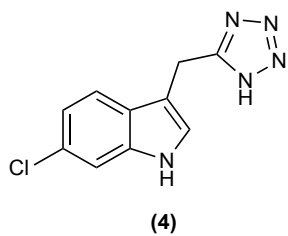

(4)

To the solution of 6-chloroindole-3-acetonitrile (75 mg, 0.39 mmol) in dimethylformamide (3mL) was added ammonium chloride (85 mg, 1.59 mmol) and sodium azide (64mg, 0.98mmol). The reaction mixture was then stirred for 28 h at 120˚C. The resulting solution was cooled and added to water (30 mL), and extracted with ethyl acetate [EtOAc] (20 mL × 2). The organic layer was washed with saturated NH$_4$Cl solution and brine, and then dried over Na$_2$SO$_4$. The residue was purified by a silica gel column chromatography (Chloroform:acetone=4:1) to give 3-((1H-tetrazol-5-yl)methyl)-6-chloro-1H-indole as white powder (38 mg, 41% yield):

$^1$H NMR (500 MHz, DMSO-d$_6$) δ = 11.15 (s., 1H), 7.43 (d, J = 8.5 Hz, 1H), 7.41 (s., 1H), 7.30 (s, 1H), 7.00 (d, J = 8.50 Hz, 1H), 4.36 (s, 2H); $^{13}$C NMR (126 MHz, DMSO-d$_6$) δ = 155.9, 136.9, 126.4, 125.7, 125.4, 119.9, 119.3, 111.5, 109.1, 19.7; HRMS (*m/z*): [M]$^-$ calcd. For C$_{10}$H$_7$ClN$_5$, 232.0395; found, 232.0396. FAB-MS (*m/z*): [M+H]$^+$ 234.

# Supporting Information Table 1:

| ID | IAA | iMT | 4-Cl-IAA | 5-Cl-IAA | 6-Cl-IAA | 4-Cl-iMT | 5-Cl-iMT | 6-Cl-iMT |
|---|---|---|---|---|---|---|---|---|
| Total Molweight | 175.186 | 199.216 | 209.632 | 209.632 | 209.632 | 233.662 | 233.662 | 233.662 |
| pKa | 4.66 | 4.84 | 4.11 | 4.11 | 4.11 | 4.84 | 4.84 | 4.84 |
| cLogP | 1.1822 | 1.0278 | 1.7882 | 1.7882 | 1.7882 | 1.6338 | 1.6338 | 1.6338 |
| H-Acceptors (pH 7.4) | 2 | 4 | 2 | 2 | 2 | 4 | 4 | 4 |
| H-Donors (pH 7.4) | 1 | 1 | 1 | 1 | 1 | 1 | 1 | 1 |
| Total Surface Area | 135.29 | 156.8 | 150.71 | 150.71 | 150.71 | 172.22 | 172.22 | 172.22 |
| Polar Surface Area | 53.09 | 70.25 | 53.09 | 53.09 | 53.09 | 70.25 | 70.25 | 70.25 |
| Rotatable Bonds | 2 | 2 | 2 | 2 | 2 | 2 | 2 | 2 |
| Aromatic Rings | 2 | 3 | 2 | 2 | 2 | 3 | 3 | 3 |
| Charge at pH 7.4 | -1 | -1 | -1 | -1 | -1 | -1 | -1 | -1 |

**Table 1: Physiochemical properties of IAA and iMT analogues**

# Supporting Information Figure 1

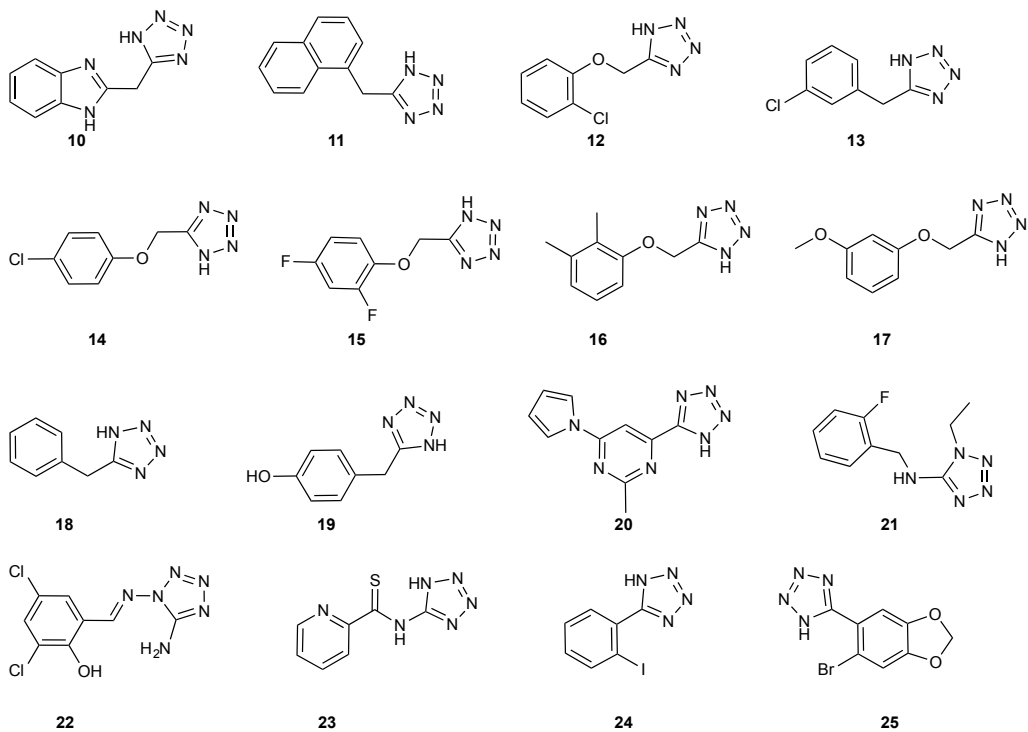

**Figure 1: Additional tetrazoles tested for auxin-like activity against TIR1 and AFB5.**

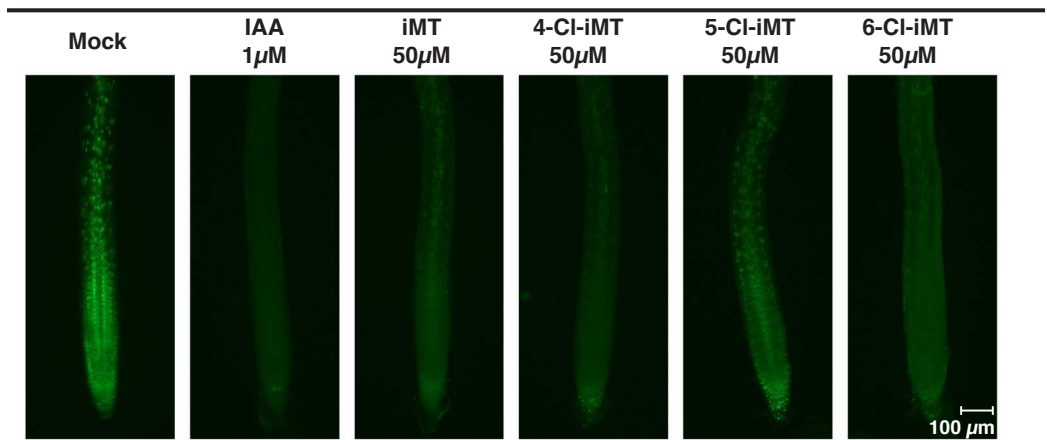

**Fig.** ... GFP signal seen ... 50μM, exce...

# Supporting Information Figure 3

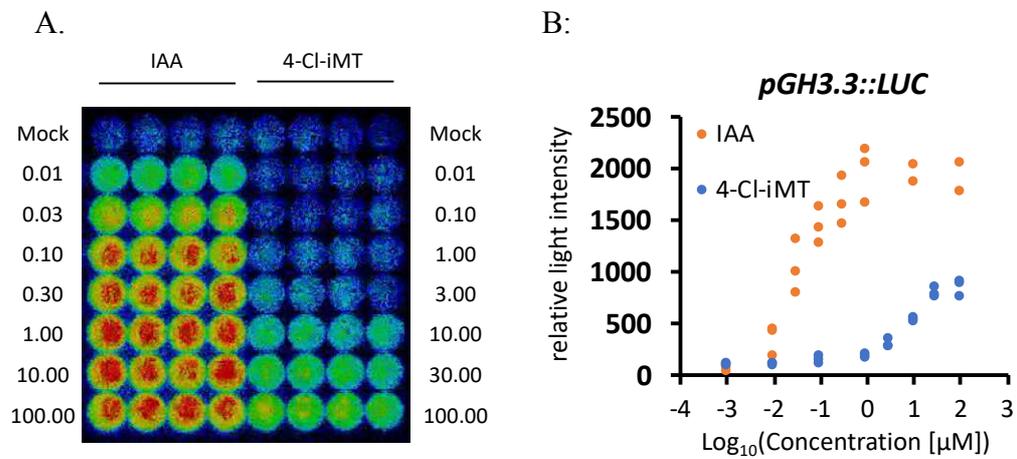

**Figure 3: Protoplast activity assays**. A: Replicate wells of protoplasts were exposed to serial dilutions of IAA (left) or 4-Cl-iMT (right). B: Luciferase readings plotted against dose indicate the differences in potency and time to response between the two compounds.

# Supporting Information Figure 4

# Root growth inhibition curves

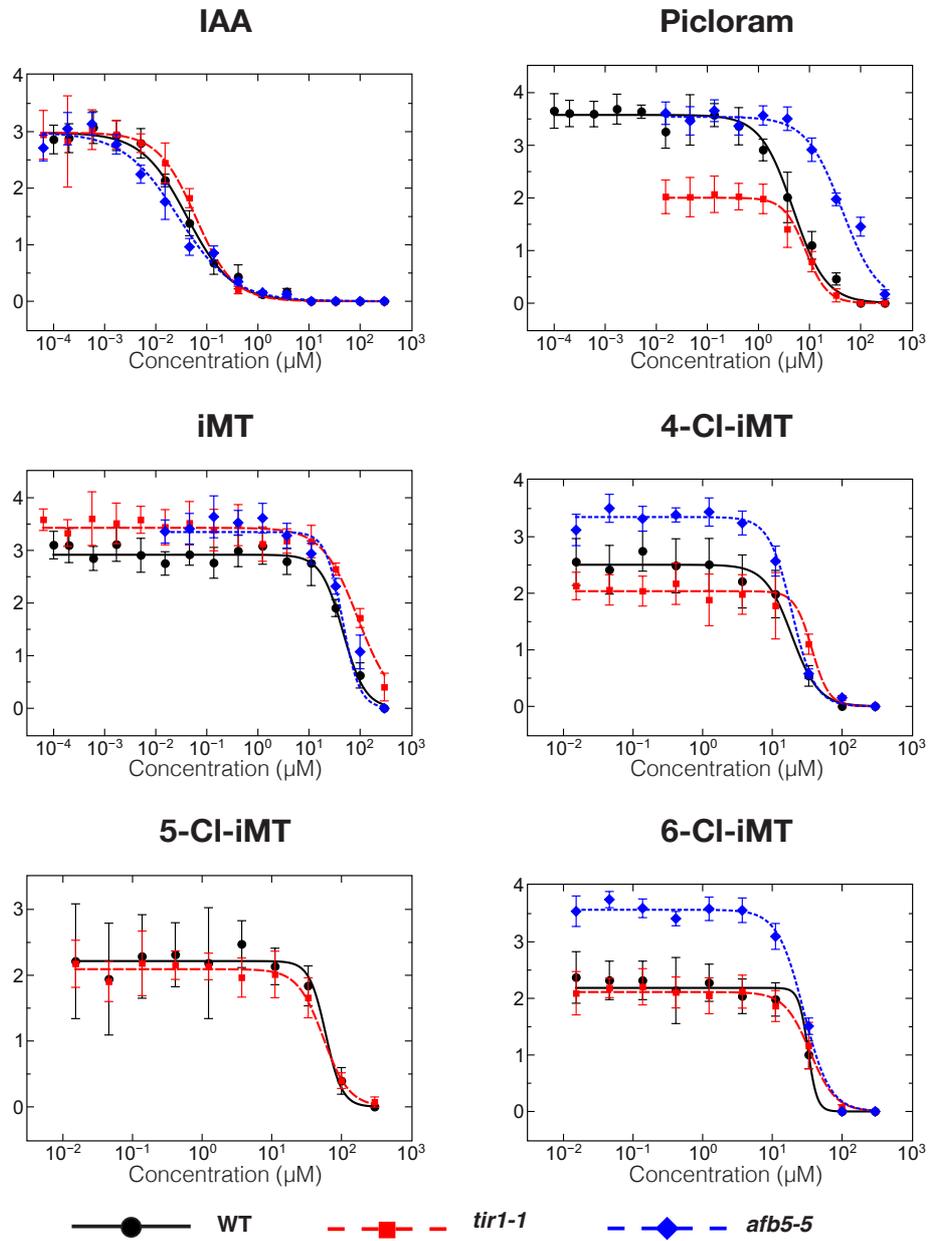

**Figure 4: Primary root growth inhibition dose response curves for compounds tested against tir1-1 and afb5-5 lines.**

Supporting Information Figure 5

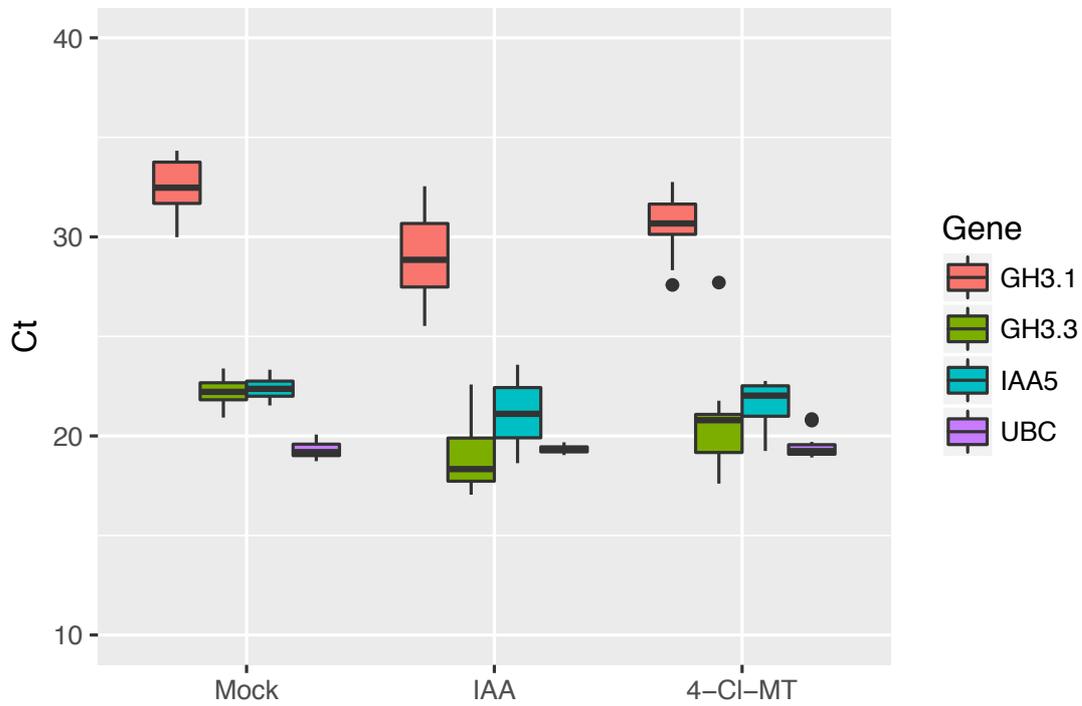

**Supplementary Figure 5**. A boxplot showing that *UBC* expression is stable across treatments and replicates. Cycle threshold ($C_T$) values are given from four biological replicates after Mock (1% DMSO), 1 µM IAA or 10 µM 4-Cl-MT treatments. For *GH3.1*, *GH3.3* and *IAA5* genes the $C_T$ value declines (expression is increased) after IAA and 4-Cl-MT treatments, whereas $C_T$ for *UBC* remains stable with little variation across samples and conditions.

# Supporting Information Table 2

Within each row, compare columns (simple effects within rows)

| | | | | | |
|---|---|---|---|---|---|
| Number of families | 10 | | | | |
| Number of comparisons per family | 2 | | | | |
| Alpha | 0.05 | | | | |

| Sidak's multiple comparisons test | Mean Diff. | 95.00% CI of diff. | Significant? | Summary | Adjusted P Value |
|---|---|---|---|---|---|
| 300 | | | | | |
| WT vs. tir1-1 | 6.8 | 5.841 to 7.759 | Yes | **** | <0.0001 |
| WT vs. afb5-5 | -1.2 | -2.159 to -0.2408 | Yes | * | 0.0105 |
| | | | | | |
| 100 | | | | | |
| WT vs. tir1-1 | 1.5 | 0.5408 to 2.459 | Yes | ** | 0.001 |
| WT vs. afb5-5 | 0.5 | -0.4592 to 1.459 | No | ns | 0.4256 |
| | | | | | |
| 33.33333 | | | | | |
| WT vs. tir1-1 | 2.2 | 1.241 to 3.159 | Yes | **** | <0.0001 |
| WT vs. afb5-5 | -0.6 | -1.559 to 0.3592 | No | ns | 0.2955 |
| | | | | | |
| 11.11 | | | | | |
| WT vs. tir1-1 | 1.3 | 0.3408 to 2.259 | Yes | ** | 0.0051 |
| WT vs. afb5-5 | 0.3 | -0.6592 to 1.259 | No | ns | 0.7321 |
| | | | | | |
| 3.703704 | | | | | |
| WT vs. tir1-1 | 0.8 | -0.1592 to 1.759 | No | ns | 0.1198 |
| WT vs. afb5-5 | -0.1 | -1.059 to 0.8592 | No | ns | 0.9657 |
| | | | | | |
| 1.234568 | | | | | |
| WT vs. tir1-1 | 0.7 | -0.2592 to 1.659 | No | ns | 0.1935 |
| WT vs. afb5-5 | -0.2 | -1.159 to 0.7592 | No | ns | 0.8701 |
| | | | | | |
| 0.4115226 | | | | | |
| WT vs. tir1-1 | 0.5 | -0.4592 to 1.459 | No | ns | 0.4256 |
| WT vs. afb5-5 | -0.5 | -1.459 to 0.4592 | No | ns | 0.4256 |
| | | | | | |
| 0.1371742 | | | | | |
| WT vs. tir1-1 | 0.2 | -0.7592 to 1.159 | No | ns | 0.8701 |
| WT vs. afb5-5 | -0.4 | -1.359 to 0.5592 | No | ns | 0.5764 |
| 0.04572474 |  |  |  |  |  |
| WT vs. tir1-1 | 1 | 0.04075 to 1.959 | Yes | * | 0.0392 |
| WT vs. afb5-5 | 0.2 | -0.7592 to 1.159 | No | ns | 0.8701 |
| 0.01524158 |  |  |  |  |  |
| WT vs. tir1-1 | 0.8 | -0.1592 to 1.759 | No | ns | 0.1198 |
| WT vs. afb5-5 | 0.4 | -0.5592 to 1.359 | No | ns | 0.5764 |

**Table 2: Two-way ANOVA test of lateral root densities, comparing *tir1-1* and *afb5-5* mutant lines to Col-0 across the dose response range of 4-Cl-iMT treatments.**

Supporting information Table 3

| Gene ref. | Gene | F/R | forward/reverse primer sequence |
|---|---|---|---|
| Reference genes: | | | |
| AT5G25760 | UBC | F | AACTGCGACTCAGGGAATCT |
| AT5G25760 | UBC | R | GCGAGGCGTGTATACATTTG |
| Test genes: | | | |
| AT2G14960 | GH3.1 | F | CTCCCATCTTATCTGCCCAT |
| AT2G14960 | GH3.1 | R | GGTCGGCATAAGTTTCCTCT |
| AT2G23170 | GH3.3 | F | GGAGATTCAACGTATTGCCA |
| AT2G23170 | GH3.3 | R | GGTTGGCATCAACTTCCTTT |
| AT1G15580 | IAA5 | F | CGTTGAAGGAAAGTGAATGTG |
| AT1G15580 | IAA5 | R | ATCCAAGGAACATTTCCCAA |